\documentclass[3p,times,twocolumn]{elsarticle}

\usepackage{ecrc}

\volume{00}

\firstpage{1}

\journalname{Nuclear Physics B Proceedings Supplement}

\runauth{}

\jid{nuphbp}

\jnltitlelogo{Nuclear Physics B Proceedings Supplement}

\usepackage{amssymb}
\usepackage{amsthm}

\usepackage{lineno}

\usepackage{mathrsfs}
\usepackage{amsmath,amssymb,dsfont,placeins}
\usepackage{verbatim}
\usepackage{stmaryrd,xcolor}
\usepackage{amsfonts,graphicx,epsfig,float,ulem}
\usepackage{bbm}

\newcommand{\mr}{\mathrm}

\usepackage[figuresright]{rotating}

\begin{document}

\begin{frontmatter}



\dochead{}

\title{Light neutrino mass spectrum with one or two right-handed singlet fermions added}


\author[itpa]{Darius Jur\v{c}iukonis}
\author[ff]{Thomas Gajdosik}
\author[itpa]{Andrius Juodagalvis}

\vspace{0.6cm}

\address[itpa]{Vilnius University, Institute of Theoretical Physics and Astronomy, A. Go\v{s}tauto st. 12,  LT-01108 Vilnius, Lithuania}

\address[ff]{Vilnius University, Physics Faculty, Saul\.{e}tekio al. 9, LT-10222 Vilnius, Lithuania}

\begin{abstract}
We analyse two cases of the minimal extension of the Standard Model when
one or two right-handed fields are added to the three left-handed
fields. A second Higgs doublet (two Higgs doublet model -- 2HDM) is included in our model. We calculate one-loop radiative corrections to the mass parameters
which produce mass terms for the neutral leptons. In both cases we
numerically analyse light neutrino masses as functions of the heavy
neutrino masses. Parameters of the model are varied to 
find light neutrino masses that are compatible with experimental data of
solar $\Delta m^2_\odot$ and atmospheric $\Delta m^2_\mathrm{atm}$
neutrino mass differences for normal hierarchy. We choose values for the parameters of the tree-level by numerical scans, where we look for the best agreement between computed and experimental neutrino oscillation angles.
\end{abstract}


\end{frontmatter}


\section{Description of the model}
\label{framework}

The mass terms for the neutrinos can be written in a compact form
as a mass term with a $(n_L+n_R) \times (n_L+n_R)$ symmetric mass matrix
\begin{equation}\label{Mneutr}
M_{\nu} = 
\left( \begin{array}{cc} 0 & M_D^T \\
M_D & \hat{M}_R \end{array} \right),
\end{equation}
where $M_D$ is a $n_L \times n_R$ Dirac neutrino mass matrix, while the hat indicates that $\hat{M}_R$ is a diagonal matrix. $M_{\nu}$ can be diagonalized as
\begin{equation}\label{Mtotal}
U^T M_{\nu}\, U = \hat m
= \mathrm{diag} \left( m_1, m_2, \ldots, m_{n_L+n_R} \right),
\end{equation}
where the $m_i$ are real and non-negative.
In order to implement the seesaw mechanism \cite{seesaw}
we assume that the elements of $M_D$ are of order $m_D$
and those of $M_R$ are of order $m_R$,
with $m_D \ll m_R$.
Then,
the neutrino masses $m_i$ with $i=1, 2, \ldots, n_L$
are of order $m_D^2/m_R$,
while those with $i = n_L+1, \ldots, n_L+n_R$ are of order $m_R$. 

In the standard seesaw, one-loop corrections to the mass matrix,
i.e.\ the self energies, are determined by the neutrino interactions
with the $Z$ boson, the neutral Goldstone boson $G^0$, and the Higgs
boson $h^0$. Each diagram contains a divergent
piece but the sum of these three contributions turns out
to be finite.

Once the one-loop corrections are taken into account
the neutral fermion mass matrix is given by
\begin{equation}\label{M1}
M^{(1)}_\nu = \left(  \renewcommand\arraycolsep{2pt} \begin{array}{cc}
\delta M_L & M_D^T+\delta M_D^T \\
M_D+\delta M_D    & \hat{M}_R+\delta M_R\end{array} \right)\approx
\left(  \renewcommand\arraycolsep{2pt} \begin{array}{cc}
\delta M_L & M_D^T \\
M_D   & \hat{M}_R\end{array} \right)
\end{equation}
where the $0_{3\times3}$ matrix appearing at tree level (\ref{Mneutr})
is replaced by the contribution $\delta M_L$. This correction is a
symmetric matrix, it has the largest influence as compared
to other corrections.

The expression for one-loop corrections is given by \cite{Grimus:2002nk} 
\begin{align}
&\delta M_L =\sum_{b} \frac{1}{32 \pi^2}\, \Delta_b^T U_R^\ast \hat m
\left(
    \frac{\hat {m}^2}{m_{H^0_b}^2}-\mathbbm{1}
  \right)^{-1}\hspace{-5pt}
  \ln\left(\frac{\hat {m}^2}{m_{H^0_b}^2}\right) U_R^\dagger \Delta_b \notag \\
& \phantom{W}+ \frac{3 g^2}{64 \pi^2 m_W^2}\, M_D^T U_R^\ast \hat m
\left(
    \frac{\hat {m}^2}{m_Z^2}-\mathbbm{1}
  \right)^{-1}\hspace{-5pt}
  \ln\left(\frac{\hat {m}^2}{m_Z^2}\right) U_R^\dagger M_D,
\end{align}
with $\Delta_b = \sum_k b_k \Delta_k$, where $b_k$ are two-dimensional
complex unit vectors, $\Delta_k$ are the Yukawa matrices,
and the sum $\sum_b$ runs over all neutral physical Higgses $H^0_b$.

Neutral Higgses are characterized by the unit-length vectors $b$ which hold special orthogonality conditions. We use the set of values
\begin{equation}
b_Z = \left( \begin{array}{c} i \\ 0 \end{array} \right), \hspace{0.02cm}  b_1 = \left( \begin{array}{c} 1 \\ 0 \end{array} \right), \hspace{0.02cm}
b_2 = \left( \begin{array}{c} 0 \\ 1 \end{array} \right), \hspace{0.02cm}  b_3 = \left( \begin{array}{c} 0 \\ i \end{array} \right).
\label{bVectSetTh}
\end{equation}
It coincides with a special case of the invariant mixing angles
that define the diagonalization matrix of the neutral Higgs
squared-mass matrix \cite{Haber11}. The Higgs potential of the 2HDM
becomes CP-conserving in this case.


\section{Case $n_R=1$}
\label{3x1}

First we consider the minimal extension of the standard model by adding only one right-handed field $\nu_R$ to the three left-handed fields contained in $\nu_L$.

We define $\Delta_i = \left(\sqrt{2}\, m_D/v \right)\vec{a}_i^T$, where  $\vec{a}_1^T = \bigl(0,0,1\bigr)$ and $\vec{a}_2^T = \bigl(0,n,e^{i \phi} \sqrt{1-n^2}\bigr)$. Diagonalization of the symmetric mass matrix $M_{\nu}$ (\ref{Mneutr}) in block form is 
\begin{equation}\label{Mneutr1}
U^{T}M_{\nu}U = 
U^{T}\left(\renewcommand{\arraystretch}{1} \renewcommand\arraycolsep{3pt} \begin{array}{cc} 0_{3\times3} & m_{D} \vec{a}_1\\
m_D \vec{a}_1^T & \hat{M}_R \end{array} \right)U=\left(\renewcommand{\arraystretch}{1}  \renewcommand\arraycolsep{3pt} \begin{array}{cc} \hat{M}_l & 0 \\
0 & \hat{M}_h \end{array} \right).
\end{equation}
The non zero masses in $\hat{M}_l$ and $\hat{M}_h$ are determined analytically by finding eigenvalues of the hermitian matrix $M_{\nu}M^{\dagger}_{\nu}$. These eigenvalues are the squares of the masses of the neutrinos $\hat{M}_l=\text{diag}(0,0,m_l)$ and $\hat{M}_h=m_h$. Solutions $m^2_D=m_hm_l$ and $m^2_R=(m_h-m_l)^2 \approx m^2_h$ correspond to the seesaw mechanism.

It is possible to estimate masses of the light neutrinos from experimental data of solar and atmospheric neutrino oscillations \cite{Forero} assuming that the lightest
neutrino has initial mass value $m_{\mr{in}}$. Numerically varying the value of $m_{\mr{in}}$ we choose parameters for the tree level that agree with the experimental oscillation data most accurately.

For the calculation of radiative corrections we use the orthogonal complex vectors $b$ defined in eq.~(\ref{bVectSetTh}). Diagonalization of the mass matrix corrected by one-loop contributions is performed with a unitary matrix $U_{\text{loop}}=U_{\text{egv}}U_{\varphi}(\varphi_1,\varphi_2,\varphi_3)$, where $U_{\text{egv}}$ is an eigenmatrix of $M^{(1)}_{\nu}M^{(1)\dagger}_{\nu}$, and $U_{\varphi}$ is a phase matrix. The second light neutrino obtains its mass from radiative corrections. The third light neutrino remains massless.

\begin{figure}[H]
\begin{center}
\includegraphics[scale=0.5]{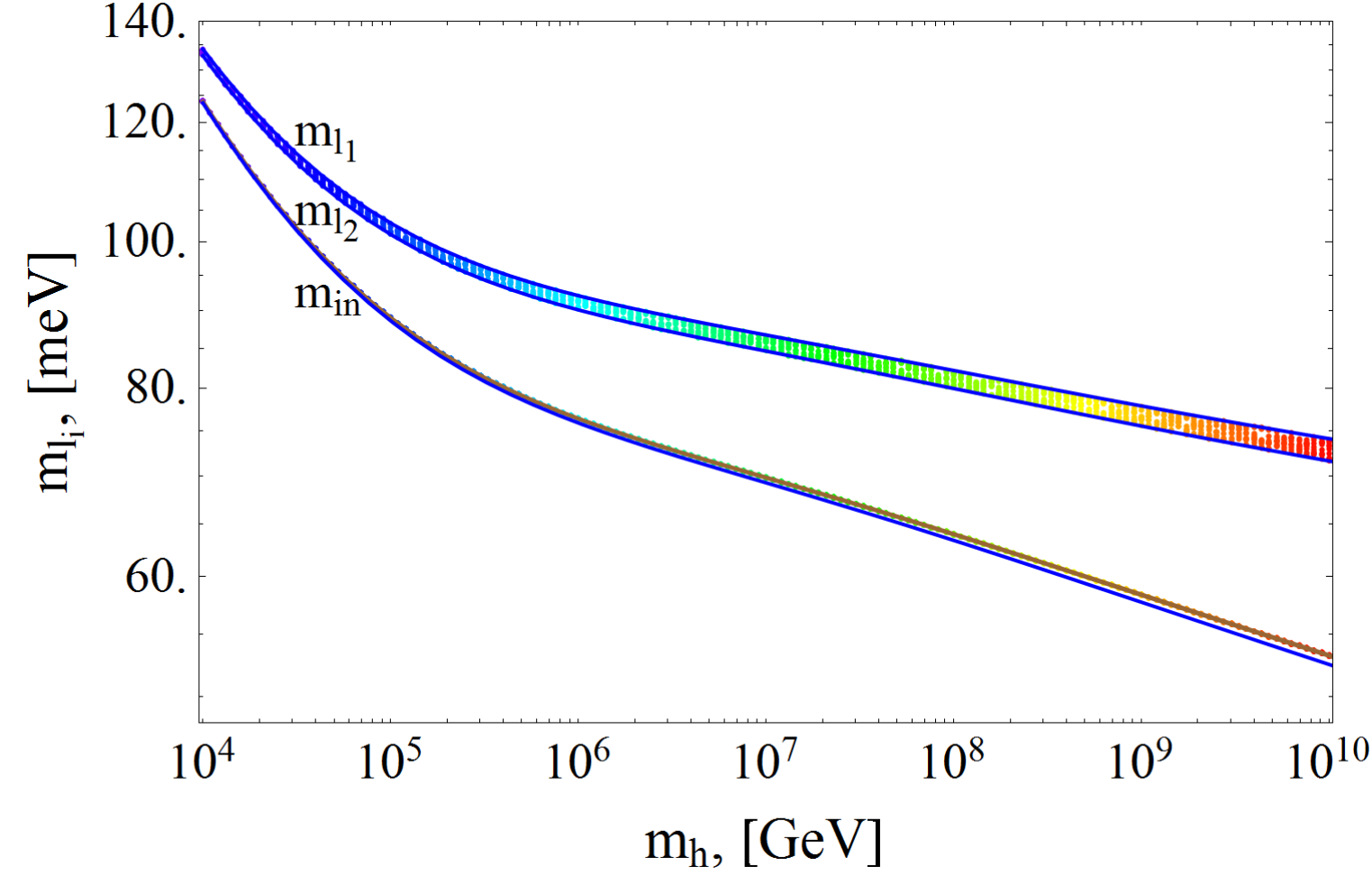}
\end{center}
\vspace{-0.6cm}
\caption{(Color online)
Calculated masses of two light neutrinos as a function of the heavy neutrino mass $m_h$.
Values of $m_{l_1}$ ($m_{l_2}$) are shown as a band (brown line).
The blue solid line  represents the value of $m_{\mr{in}}$ used for the calculations of the tree level masses.}
\label{picture1}
\end{figure}

\begin{figure}[H]
\begin{center}
\includegraphics[scale=0.5]{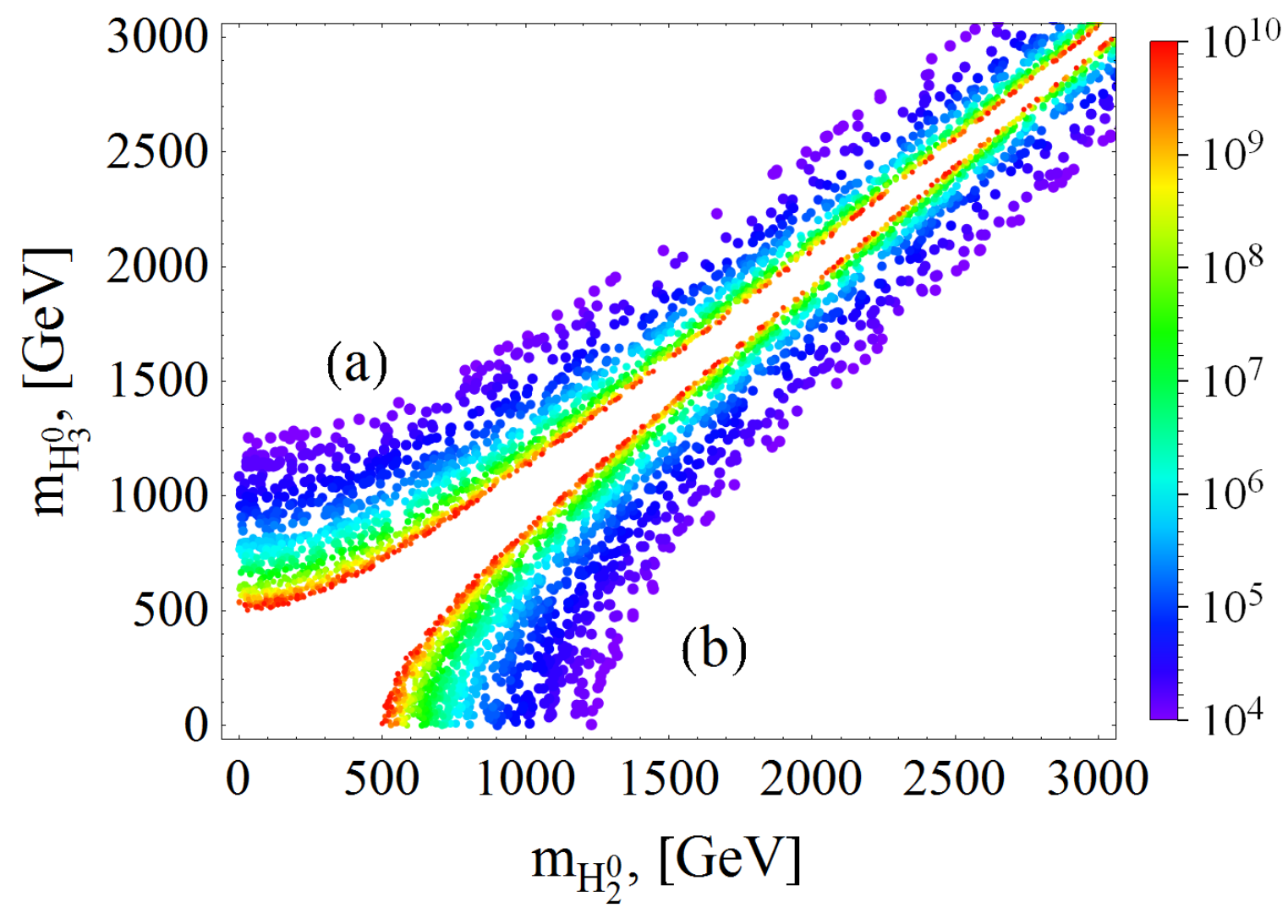}
\end{center}
\vspace{-0.6cm}
\caption{(Color online)
The values of the free parameters $m_{H^0_2}$ and $m_{H^0_3}$ as functions of the heaviest right-handed neutrino mass $m_{h}$.
The scale of the $m_h$ values is shown on the right.
The mass of the SM Higgs boson is fixed to $m_{H^0_1}=125$~GeV.}
\label{picture2}
\end{figure}

The numerical analysis shows (see Fig.~\ref{picture1}) that we can reach the allowed neutrino mass ranges for a heavy singlet with the mass close to $10^{4}$~GeV when the angle of oscillations $\theta_{\mathrm{atm}}$ is fixed to the experimental $3\sigma$ range  \cite{Forero}.
The free parameters $n$, $\phi$, $m_{H^0_2}$ and $m_{H^0_3}$ are restricted by the parametrization used and by the oscillation data. Figure \ref{picture2} illustrates the  allowed values of Higgs masses for different values of the heavy singlet. The values of Higgs masses spread to two separated sets.

\section{Case $n_R=2$}
\label{3x2}

When we add two singlet fields $\nu_R$ to the three left-handed fields
$\nu_L$, the radiative corrections give masses to all three light
neutrinos.

We parametrize $\Delta_i=\frac{\sqrt{2}}{v}\,\left(m_{D_2} \vec{a}_i^T , m_{D_1} \vec{b}_i^T  \right)^T$ with $|\vec{a}|=1$, and $|\vec{b}|=1$. Diagonalizing the symmetric mass matrix $M_{\nu}$ (\ref{Mneutr}) in block form we write: 
\begin{equation}\label{Mneutr2}
U^{T}M_{\nu}U =\!
U^{T}\!\left(\renewcommand{\arraystretch}{1}  \renewcommand\arraycolsep{.89pt}\begin{array}{cc} 0_{3 \times 3} & m_{D_2} \vec{a} \hspace{0.09cm} m_{D_1} \vec{b}\\
\begin{array}{c} m_{D_2} \vec{a}^T \\ m_{D_1} \vec{b}^T \end{array} & \hat{M}_R \end{array} \right)\hspace{-0.03cm}U \hspace{-0.05cm} =\left(  \renewcommand\arraycolsep{1pt} \begin{array}{cc} \hat{M}_l & 0 \\
0 & \hat{M}_h \end{array} \right) \hspace{-0.06cm}.
\end{equation}
The non zero masses in $\hat{M}_l$ and $\hat{M}_h$ are determined by
the seesaw mechanism: $m^2_{D_i}\approx m_{h_i}m_{l_i}$ and $m^2_{R_i}
\approx m^2_{h_i}$, $i=1,2$. We use $m_1>m_2>m_3$ ordering
of masses. At tree level the third light neutrino is massless.

\begin{figure}[H]
\begin{center}
\includegraphics[scale=0.5]{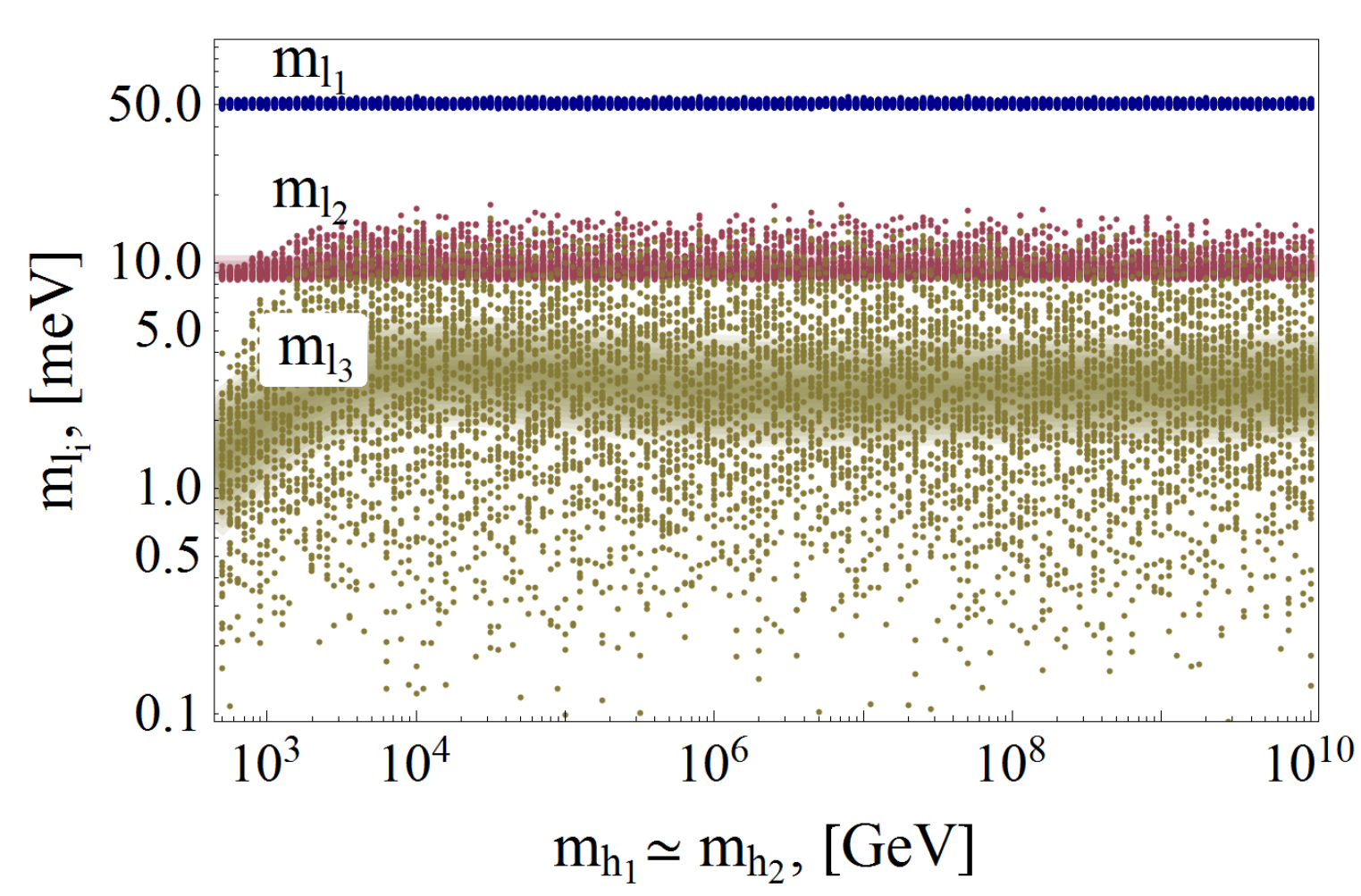}
\end{center}
\vspace{-0.6cm}
\caption{(Color online)
The masses $m_{l_i}$ of the light neutrinos as functions of
  the heaviest right-handed neutrino mass $m_{h_{1,2}}$ for the normal
  hierarchy of the light neutrinos. The wide solid lines indicate the area of the
  most frequent values of the scatter data.}
\label{picture3}
\end{figure}

\begin{figure}[H]
\begin{center}
\includegraphics[scale=0.5]{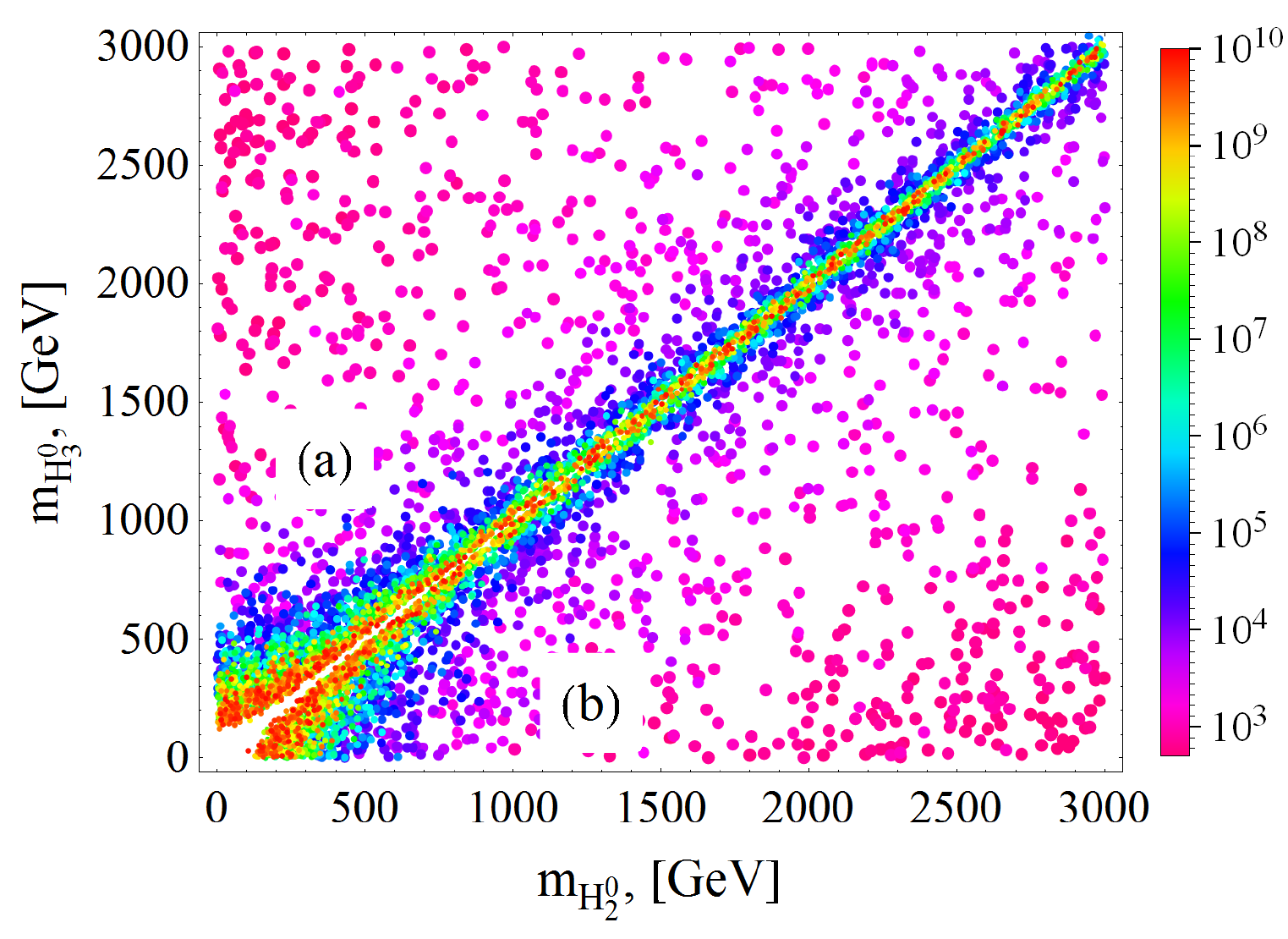}
\end{center}
\vspace{-0.6cm}
\caption{(Color online)
The values of the free parameters $m_{H^0_2}$ and $m_{H^0_3}$ as functions of the heaviest right-handed neutrino mass $m_{h_{1,2}}$ for the normal hierarchy of the light neutrinos.}
\label{picture4}
\end{figure}

In numerical calculations the model parameters as well as the derived
masses of the light neutrinos are obtained in several steps.  First,
the diagonal mass matrix for the tree level is constructed. Having chosen the mass for the third neutrino $m_{\mr{in}}$, the masses of the other two light neutrinos are estimated from experimental data on solar and atmospheric neutrino oscillations \cite{Forero}. The masses of the heavy right-handed neutrinos are input
parameters. This diagonal tree level matrix is used to constrain the parameters
that enter the tree-level mass matrix $M_\nu$
and its diagonalization matrix. Then the diagonalization matrix is used to
evaluate one-loop corrections to the mass matrix. Diagonalization of
the corrected mass matrix yields masses for the three light neutrinos. If the
calculated mass difference is compatible with the experimental oscillations data, the parameter set is kept. Otherwise,
another set of parameters is generated. Figure \ref{picture3} illustrates the obtained results for normal neutrino mass ordering.
Figure \ref{picture4} illustrates the  allowed values of Higgs masses for different values of the heavy singlet. The values of Higgs masses spread to two separated sets as for the $n_R=1$ case.
In the case of $n_R = 2$ the inverted hierarchy is also allowed
\cite{Jurciukonis}.

\section{Conclusions} \label{concl}

\begin{enumerate}
\item For the case $n_R=1$ we can match the differences of the calculated
light neutrino masses to $\Delta m^2_\odot$ and $\Delta
m^2_\mathrm{atm}$ with the mass of a heavy singlet larger than
$10^4$~GeV. The parametrization used for this case and restrictions from the neutrino oscillation data limit the values of free parameters.
One light neutrino remains massless.

\item In the case $n_R=2$ we obtain three non vanishing masses. The numerical analysis shows that the values of parameters and Higgs masses depend on the choice of the set of $b$ vectors and the heavy neutrino masses. The
radiative corrections generate the lightest neutrino mass and have a
big impact on the second lightest neutrino mass.

\end{enumerate}

\noindent {\bf Acknowledgement}
 
The authors thank the Lithuanian Academy of Sciences for the support (the project DaFi2014, No.\ CERN-VU-2014-1).




\nocite{*}
\bibliographystyle{elsarticle-num}
\bibliography{martin}







\end{document}